\documentclass[sigconf,manuscript]{acmart}

\AtBeginDocument{%
  \providecommand\BibTeX{{%
    \normalfont B\kern-0.5em{\scshape i\kern-0.25em b}\kern-0.8em\TeX}}}

\setcopyright{acmcopyright}
\copyrightyear{2021}
\acmYear{2021}

\acmConference[CHI '21]{CHI '21 Workshop: Realizing AI in Healthcare: Challenges Appearing in the Wild}{May 08--09, 2021}{Yokohama, Japan}

\usepackage{booktabs}
\usepackage{textcomp}
\usepackage{enumitem}
\usepackage{fixme}
\fxsetup{status=draft, layout=inline, theme=color}
\usepackage{framed}

\FXRegisterAuthor{ch}{ach}{cait}
\FXRegisterAuthor{gm}{agm}{gabi}
\FXRegisterAuthor{fc}{afc}{fanny}
\FXRegisterAuthor{ag}{aag}{anna}

\begin{document}


\title{Considerations for Visualizing Uncertainty in Clinical Machine Learning Models}

\author{Caitlin F. Harrigan}
\authornote{Both authors contributed equally to this research.}
\email{cait.harrigan@mail.utoronto.ca}
\affiliation{%
  \institution{Department of Computer Science, University of Toronto}
  \city{Toronto}
  \country{Canada}
}
\affiliation{%
  \institution{Vector Institute}
  \city{Toronto}
  \country{Canada}} 
\orcid{1234-5678-9012}

\author{Gabriela Morgenshtern}
\authornotemark[1]
\email{morgensh@cs.toronto.edu}
\affiliation{%
  \institution{Department of Computer Science, University of Toronto}
  \city{Toronto}
  \country{Canada}
}
\affiliation{%
  \institution{Genetics and Genome Biology, The Hospital for Sick Children}
  \city{Toronto}
  \country{Canada}}
\affiliation{%
  \institution{Vector Institute}
  \city{Toronto}
  \country{Canada}} 
\orcid{0000-0003-4762-8797}

\author{Anna Goldenberg}
\email{anna.goldenberg@utoronto.ca}
\affiliation{%
  \institution{Department of Computer Science, University of Toronto}
  \city{Toronto}
  \country{Canada}
}
\affiliation{%
  \institution{Genetics and Genome Biology, The Hospital for Sick Children}
  \city{Toronto}
  \country{Canada}}
\affiliation{%
  \institution{Vector Institute}
  \city{Toronto}
  \country{Canada}} 

\author{Fanny Chevalier}
\email{fanny@cs.toronto.edu}
\affiliation{%
  \institution{Department of Computer Science, University of Toronto}
  \city{Toronto}
  \country{Canada}
}
\renewcommand{\shortauthors}{Harrigan  and Morgenshtern, et al.}

\begin{abstract}

Clinician-facing predictive models are increasingly present in the healthcare setting. Regardless of their success with respect to performance metrics, all models have uncertainty. We investigate how to visually communicate uncertainty in this setting in an actionable, trustworthy way. To this end, we conduct a qualitative study with cardiac critical care clinicians. Our results reveal that clinician trust may be impacted most not by the degree of uncertainty, but rather by how transparent the visualization of what the \textit{sources} of uncertainty are. Our results show a clear connection between feature interpretability and clinical actionability. 

\end{abstract}

\ccsdesc[500]{Human-centered computing~Empirical studies in visualization}

\keywords{}


\begin{teaserfigure}
  \includegraphics[width=\textwidth]{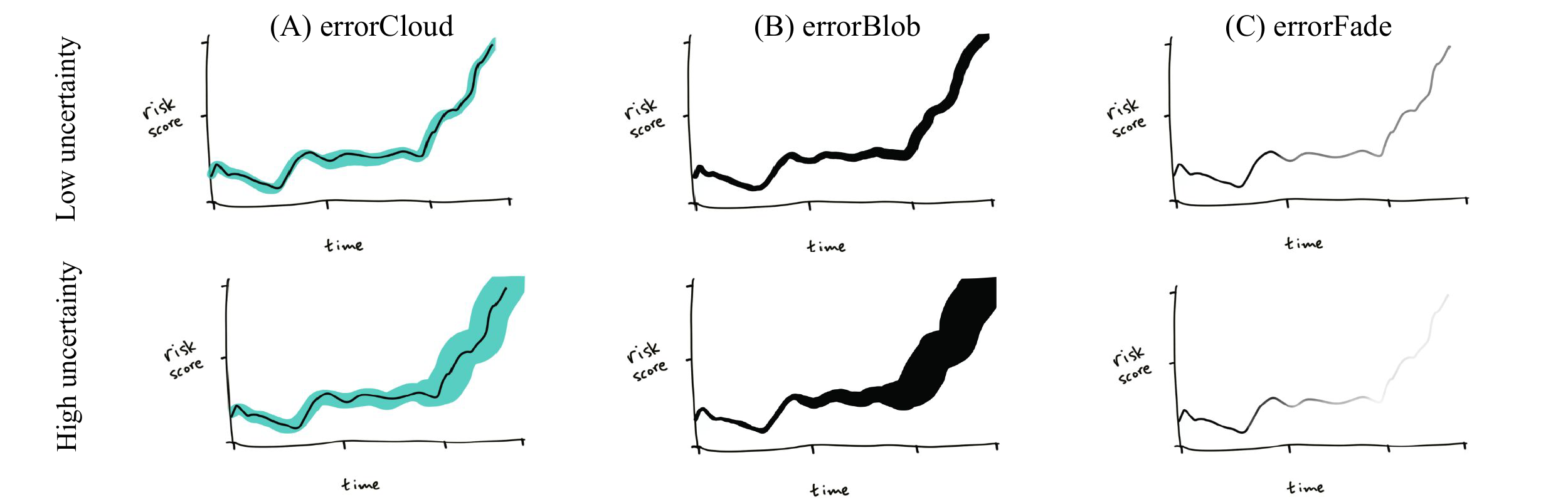}
  \vspace*{-2em}
  \caption{Sketch renderings of three ways of displaying uncertainty, used to probe clinicians in our study. A: errorCloud acts as a 'baseline'. B: errorBlob highlights areas of low uncertainty through variation on the spatial channel \cite{sanyal_user_2009}. C: errorFade highlights areas of high uncertainty through variation on colour \cite{sanyal_user_2009}. Axis scales are intentionally unmarked; relevant scale is probed on in interviews.}
  \label{fig:sketches}
\end{teaserfigure}

\maketitle

\section{Introduction}

Supporting clinical care through integrating predictive machine learning (ML) into clinician workflows has potential to improve the standard of care for patients. A predictive model is one that uses statistical approaches to generate predictions of unseen outcomes \cite{wang_big_2018}. Regardless of how robust they are, these models have uncertainty, which hinders adoption due to lack of trust~\cite{asan2020artificial}. In this work, we investigate what design considerations are perceived to most impact trust and clinical actionability when communicating predictive uncertainty, through a qualitative study.




Clinicians in the critical care unit are adept at establishing a holistic picture of patient state by mentally integrating bedside data with information derived from physical exams, patient histories, and lab results. Critical care is a particular setting, because in it the clinicians consume raw features alongside model output. A model's output is just one more data point whose uncertainty the clinician must account for.


ML models have two main types of uncertainty: noise in the data, and systematic uncertainty in the model. A deployed model must, additionally, deal with missing data, which may be missing at random, or (much more likely) missing because of some clinical complication. Accounting for uncertainty in measures and predictions is a key part of clinical reasoning on the part of the healthcare team \cite{kim_understanding_2018}. While there exists literature on visualizing uncertainty~\cite{hullman_pursuit_2019}, how such approaches, or what characteristics of uncertainty may affect trust and actionability in clinical practice, is poorly understood. This work aims to fill that gap. We conducted interviews with 5 clinicians to understand: 1) how clinicians' perception of uncertainty impacts trust and actionability; 2) what barriers exist in making ML predictions amenable to clinical inference; 3) how these insights can inform visualization design. We take a model of cardiac arrest as a case study, but aspects of our findings may be generalizable to visualizations in other patient care environments.

We define a clinically \textbf{actionable} visual as one which has the potential to inform clinical decision making. For example, increasing the frequency of patient bedside or remote monitoring. \textbf{Trust} is the level of perceived credibility attributed to a visualization. In ML literature, the degree of trustworthiness in a model results is strongly related to its interpretability \cite{guo_calibration_2017}. Our clinician interviews suggest that trust and visualization actionability are most positively impacted when design prioritizes transparent communication around missing data and the overall prediction trend.





\section{Background}

\subsection{Related Work}\label{sec:related work}

Our work is similar to that of Jeffery et al. \cite{jeffery_participatory_2017}, who employ participatory design strategies to explore nurses’ preferences for the display of a predictive model of cardiac arrest. Findings for desired visualization elements are closely aligned with our own findings, and included a temporal trendline of predicted cardiopulmonary arrest probability with an overlapping view of relevant lab values, vital signs, treatments, interventions, and a patient baseline. However, Jeffery et al. do not investigate the implications of displaying uncertainty alongside predicted values. 


Hullman's \cite{hullman_pursuit_2019} review of uncertainty visualization user studies reveals that in most evaluations concerning interpretation of uncertainty, there exists a bias in the instrumentation towards evaluating accuracy, rather than decision quality. Thus, we follow recommendations that evaluators focus on collecting participant feedback on how judgement is made, and what information they found helpful in making it \cite{hullman_pursuit_2019}. 

Kale et al. \cite{kale_visual_2020} found that when a visualization emphasizes the mean, users may become biased towards discounting associated uncertainty when making decisions. We wanted insight as to whether similar effects may be present in the clinician user group, leading us to the errorCloud representation of uncertainty in our study's design exercise (Fig.~\ref{fig:sketches}-B). Uncertainty visualization techniques explored here (Fig.~\ref{fig:sketches}) also draw from prior findings which indicate that error bars have overall poor performance, both in task accuracy and time to completion \cite{sanyal_user_2009, correll_error_2014}. Encoding uncertainty through glyph size yielded good performance when searching for data of low uncertainty, while glyph color outperformed glyph size when searching for data of high uncertainty \cite{sanyal_user_2009}. Lastly, our design exercise is prepared with a hand-drawn aesthetic, due to findings that user engagement in visualization critique is encouraged by this style \cite{wood_sketchy_2012}.

\subsection{Case Study}
Our work uses as a case study an ML model, not yet clinically deployed, which takes patient vitals as input, and continuously outputs a calculated \textit{risk score}: the probability of cardiac arrest occurring in the next 10 minutes \cite{tonekaboni_prediction_2018}. Data for prediction is collected from a bedside monitoring tool and includes heart rate, respiratory rate, pulse rate, oxygen saturation level, arterial blood pressure, and standard deviation of beat-to-beat intervals.


\section{Methods}\label{sec:methods}
Our interview guide outlines a semi-structured interview, split into two components: exploration of participants' clinical workflow and trust in predictive models; and a design reflection task, where participants are presented with visualization probes (Fig.~\ref{fig:sketches}). The visual representations are intentionally made low-fidelity, to encourage critique from participants on content, not interface, decisions \cite{wood_sketchy_2012}. In particular, the intention of these probes is not to decide which is the ``best'' sketch. Rather, we wish to prompt clinicians to reflect on different ways of interpreting uncertainty, and the potential implications these have for clinical actionability. 

We conducted expert reviews on the study instruments themselves. We selected three experts, representing different areas of interest: design research in a clinical context, clinical machine learning, and clinical care. Each expert was interviewed in a semi-structured 45-minute session, and the guide was revised to reflect the expert's feedback. This interview guide can be found in Appendix \ref{sec:interview guide}. 

We pre-registered\footnote{https://doi.org/10.17605/OSF.IO/869NS} our methodology and several predictions prior to conducting interviews. Interview data were compiled, and all pre-registered predictions were used as labels for response coding (topics (W) and (D) in Appendix \ref{sec:response coding}). In addition to these, common themes in communication (C) and end-product features (F) were selected \textit{post-hoc}. The study was performed at a local pediatric hospital, and five clinicians (2F/3M) were selected from a convenience sample. Four of five clinicians had \textgreater 10 years experience working with patients, and self reported experience with model uncertainty ranged from None (1), Beginner (1), Intermediate (2) to Expert (1). Each interview lasted between 30-40 minutes and was audio-recorded. Institutional Review Board approval was obtained prior to all data collection for this study. 

\section{Results} \label{sec:results}




Sources of uncertainty identified by clinicians include model uncertainty and data sparsity. Although displaying uncertainty was found to positively impact trust in a given score, it did not impact trust in the trend of predictions, nor the actionability of the information. Spatial encoding of uncertainty was a more trusted representation than colour.

\textbf{High uncertainty does not equal low actionability.}
All clinicians mentioned both data sparsity and model uncertainty as factors they feel impact uncertainty in a predictive model.
They reported that seeing a rise in predicted risk, whether sudden or over an extended time frame, would always prompt them to reassess the patient: \textit{``...regardless of the uncertainty around the score, when the trend is changing you'd always need to verify. Regardless of how much you don't trust the output, you'd still check in more frequently when [rising predicted risk] occurs''} (P4). 

All clinicians stressed the importance of understanding which raw data contributed to a change in risk, and a change in uncertainty. Gaps in data collection (data sparsity) were stated by all clinicians to be a circumstance which is both commonplace, and accounted-for in their decision-making.

\textbf{Clinicians mistrust statistical model prediction, still actively monitor it}.
Four of the five clinicians noted that they are less inclined to consider the outputs of a model when they are aware of data collection issues, such as when a lead has fallen off of the patient, or the patient has not been monitored long enough to collect sufficient data. Clinicians cited limitations in model's ``clinical acumen'' as a reason for low trust in predictions. Three of the five clinicians described relying on their ``gut feeling'' to assess patient state: an integration of labs, vitals, patient ``colour'' and amount of sweat. Despite this, four clinicians described actively monitoring the trend of predictions to inform their overall `mental model' of the patient condition. Clinicians reported that continuous reporting encourages them to check predictions at the bedside more often: \textit{``I would look at [this prediction] every time I walk by, if it's a continuous display''}.





\textbf{errorBlob forces a conversation around the trend's associated uncertainty.} 
All the clinicians first preferred the errorCloud (Fig.~\ref{fig:sketches}) representation of predicted risk, as it allowed them to know ``what the value is and what the error is''. Both errorBlob and errorFade were described as unintuitive by two clinicians. However, when discussing the potential impact of such a visualization on their intervention plan, errorCloud mainly prompted discussion of the predicted score alone, whereas errorBlob prompted discussion of the relationship between the score and the uncertainty trend in all clinicians. All the clinicians expressed mistrust of the errorFade representation's predictions, due to the fact that the variation in colour reminds them of looking at a faulty monitor.


\section{Discussion}\label{sec:discussion}
As in many settings, when it comes to integrating ML into everyday care, the design challenge is directly tied to the modeling challenge. We discus the characteristics of a model's output that are necessary for informing users on patient state, and lay out some design considerations for making that communication effective.

\subsection{``Explainable'' Uncertainty} \label{sec:explainable uncertainty}

Our results reveal that clinician trust may be impacted most not by a model that effectively displays its uncertainty, but by how communicative the visualizations of the \textit{sources} of uncertainty are. We anticipated that clinicians would only express concern about structural model uncertainty, and as such, that their trust in a prediction would fluctuate based on the degree of model uncertainty. Instead, results informed us that clinicians are acutely aware of data sparsity, and that trust remains low regardless of the uncertainty associated with the prediction. Designing a treatment plan is a process of mentally synthesizing data from a wide range of sources and searching for abnormalities in them that have implications for interpreting the patient's current state, and future trajectory. In this setting, clinicians consume raw input: there are few features, and they are naturally explainable. Clinicians ``verify'' model output against their own intuitive processing of the data. Thus, allowing feature querying at clinically abnormal datapoints, informing on feature importances and sparsities alongside the model output may be the most effective ways to bolster clinician trust. 

We further suggest that confidence calibration in models is not sufficient to address these concerns. A perfectly calibrated model is one that outputs a probability level that is representative of the true model likelihood \cite{guo_calibration_2017}. While model calibration is informative when interpreting a prediction, it does not achieve explainability of modelling error with respect to the features. Our results indicate that a trustworthy model is one that has high uncertainty when the clinician does.

\subsection{From ML Predictions to Clinical Inference} \label{sec:clinical inference}

A clinician's mental model is highly contextualized per-patient, and may integrate data sources which are impossible for a clinically deployed model to consider, such as relying on one's "gut feelings" (Sec. \ref{sec:results}). Thus, a model that communicates its uncertainty in terms of its features is immediately more actionable\,---\,so long as the features are interpretable. A clinician can use this as a starting point for what vitals should be monitored most closely, or even make use of this information when deciding when to discount a prediction. There are plenty of clinical attributes that a model can not reason about, but an informed clinician can. 

Our results suggest that, in the critical care setting, every patient presents as an exceptional case; the vitals related to their condition will have variable baselines. Especially in patients with chronic conditions, alerts on deviation from the clinically standard baseline lose their actionability. Clinicians indicate that an actionable model is one that highlights important data to be used in clinical inference, rather than one that attempts to do the inference itself.



\subsection{Implications for Visualization}
Visualizations designed with an emphasis on interpretability are well established as must-haves for clinician-facing ML models \cite{abdul_trends_2018}. We have discussed how this might be achieved via increased transparency in which aspects of a case are driving model uncertainty. Our results emphasize the necessity of feature interpretability research for clinical applications, in establishing the trust of expert users who interact directly with raw features. Additionally, because much of the clinician workflow is based on triaging each patient's likely trajectory, relaying predictive uncertainty may be less immediately actionable than relaying changes in patient status. We suggest the following considerations for the visualization of predictive critical-care models with interpretable features: (1) visualization of the predictions' trend over time, to give context to a specific score; and (2) top-level accessibility of information on the sources of the uncertainty, and their relative contribution to the score, to improve trust and actionability.


\subsection{The Last Mile}
Although black-box algorithms may deliver high success in performance metrics, validating utility once deployed is a distinct hurdle. Ultimately, treatment decisions come down to the team of clinicians and nurses assigned to a patient, who may or may not decide to make use of predictions. Most of all, we foresee clinically deployed models succeeding by bringing ML towards ``a senior resident in your pocket''. This will be achieved by delivering relevant information in a timely and interpretable way. Our results elucidate two possible avenues for future work in ML model visualization for supporting clinical decision-making in a critical-care context: 
\begin{enumerate}
    \item How to transparently define and communicate feature abnormalities driving model uncertainty, and the effect that low quality data collection has on predictions.
    \item Explicit support for data integration in conjunction with model predictions, potentially with awareness of the patient baseline. 
\end{enumerate}

\section{Conclusions}
This paper discusses initial results of a larger ongoing effort: we intend to conduct sessions with nurses to better inform our understanding of these results with the viewpoints from other concerned healthcare providers.

Inevitably, model and clinician will disagree at times, but just like how clinicians may disagree, this in itself is not necessarily undesirable. Effective visualization and interaction allows good models to be useful even when they are wrong, by enabling clinicians to easily incorporate model output into their mental model of a patient's trajectory. We find that the actionability of a prediction is inextricably linked to the perceived trustworthiness of the model's visualization.

\begin{acks}
We thank our expert reviewers: Nicole Sultanum, Sana Tonekaboni, and Dr. John Edmeston who provided helpful comments. Thanks also to all the study volunteers for their time and thoughtful discussion. 
\end{acks}

\bibliographystyle{ACM-Reference-Format}
\bibliography{refs}

 
\appendix
\section{Interview Guide} \label{sec:interview guide}
The following is our interview script, with notes for the investigator. Slides with sketches from Fig.~\ref{fig:sketches} were presented to the participant.


\begin{framed}

\noindent \textbf{Intro} (1-2min) --- \textit{Contextualization and consent for data collection}

We are investigating how clinicians working in critical care units interpret the uncertainty scores of predictions made by machine learning models. We are interested in hearing specifically about your personal experience with these in patient care, especially in cases where these might have somehow altered your patient care plan (whether for better or for worse). 

If it is okay with you, we would like to record this call, so that we may use it for notetaking an analysis after this session. This recording will only be shared between our research collaborators, and all quotes and notes we use in future publishing will be anonymized. Is this okay? \textit{(if yes, start recording)}

\vspace*{1em}

\textbf{Workflow} (8-10min) --- \textit{Questions about the participant's workflow, their background with machine learning, and their degree of trust in the underlying statistics}

\begin{itemize}[noitemsep]
\item Please describe your ICU patient rounds, especially those patients at risk for cardiac events.
\item Based on one of your recent cases, can you tell me what patient information was monitored to inform patient care?
\begin{itemize}[noitemsep]
\item Where specifically did you access this information?
\item In which recent cases, if any, did you consult a predicted value or a machine learning model to inform patient care?
\item Has there been a time where you did not trust the prediction, or disregarded the prediction?
\end{itemize}

\item How do you communicate changes in patient status to other clinicians?
\begin{itemize}[noitemsep]
\item \textit{(if needed)} What system exists to alert you of changes in status? 
\item How is this alert conveyed?
\end{itemize}
\end{itemize}

\vspace*{1.5em}

\textbf{Thinking exercise} (10-12 min) --- \textit{Additional probes specific to design. }

You have a patient at risk for a cardiac event, and a clinical machine learning model that outputs a risk score for the patient. This score describes their probability of a cardiac event in the next 10 minutes. 

\begin{itemize}[noitemsep]
\item What information would you like to see in a visualization of this model? 
\item How often would you check a screen that shows this model’s prediction?
\item Probe: The risk score output by this model has uncertainty associated with it, which can fluctuate. 
\begin{itemize}
    \item Describe how you interpret the uncertainty in a risk score? 
    \item Describe what you imagine are potential sources of uncertainty?
\end{itemize}
\item Probe: Over what time period do you consider relevant when assessing cardiac arrest risk? 
\begin{itemize}[noitemsep]
\item When would such a prediction draw your attention?
\item What would draw your attention more: a steadily increasing risk or a sudden increase in risk?
\item If the sudden increase in risk prediction had a large level of associated uncertainty, how would this affect your decision making?
\end{itemize}
\end{itemize}

Here’s a sketch representation of several approaches to displaying uncertainty of a risk score. We show three different ways of drawing two scenarios: one with low uncertainty and one with high uncertainty. Please think aloud as you look at this slide and tell us about your initial impressions, gut feelings you would like to share. \textit{(Get overall initial impression, then repeat for each column in the table separately)}

\begin{itemize}[noitemsep]
\item What information stands out to you in this visualization?
\item What information is missing for making this visualization actionable?
\item How would seeing such a visualization impact your intervention plan?
\item What do you think each visualization is best suited for? 
\item What do you still want to know?
\end{itemize}

\vspace*{2em}
\textbf{Feedback / closing} (3-5min)
Thank you for taking the time to chat with us! Do you have any last thoughts or comments for this study or the researchers?

\end{framed}

\section{List of Labels Used In Response Coding} \label{sec:response coding}

\begin{table}[H]
    
    \small
  \caption{Coding Labels. Concepts in Workflow (W) and Design (D) are derived from pre-registered predictions. Concepts in Communication (C) and Feature Descriptions (D) were generated \textit{post-hoc}.}
  \vspace*{-1em}
  \begin{tabular}{lcp{12cm}} 
    \toprule
    Topic & Concept Code & Description\\
    \midrule
    Workflow (W)  
                        &  W1  & Participants will provide an example of a time when they did not trust a statistical model. \textit{Motivation}: Modeling and deployment efforts are subject to the same weaknesses as a classical hospital test; reliability may be questionable.\\
                        &  W2  & Participants will indicate experience in using machine learning models to inform patient care. \textit{Motivation}: Participants have been selected though convenience sampling of collaborators at a local hospital who have been previously exposed to clinical machine learning modeling efforts  \\
    \addlinespace
    Design (D)   
                        &  D1  &  Participants will describe uncertainty in terms of a confidence interval. \textit{Motivation}: Basic statistical training includes confidence intervals, many hospital tests are amenable to this interpretation. \\
                        &  D1.1  &  Participants will not refer to data sparsity as a source of uncertainty \textit{Motivation}: Participants will be primed to exclusively describing uncertainty that corresponds to model inaccuracies \\
                        &  D2  &  Participants will express high valuation of risk scores with low uncertainty \textit{Motivation}: A more certain score is viewed as more actionable \\
                        &  D2.1  &  Participants will express desire to query values of model features. \textit{Motivation}: It has been previously observed in the literature that thorough understanding of a model’s results improves clinician trust. \\
                        &  D2.2  &  Participants will not indicate that rate of change is an important factor for determining their trust in a predicted value. \textit{Motivation}: The actionability of a score fluctuation depends on the size of the confidence interval, which is what chiefly determines trust in predicted value. \\
                        &  D3  & Participants will indicate familiarity with the errorCloud representation. \textit{Motivation}: Basic statistical training includes confidence intervals  \\
                        &  D4  & Participants will indicate preference for the errorBlob representation over the other two viz settings \textit{Motivation}: high visual impact of the increased visual real estate better leverages preattentive processing of users  \\
                        &  D5  &  Participants will indicate unfamiliarity with the errorFade representation. \textit{Motivation}: errorFade is not regularly employed in healthcare visualizations, does not explicitly encode an interval \\
                        &  D5.1  & Participants will indicate mistrust of values represented with errorFade \textit{Motivation}: colour channel does not leverage preattentive cognition as readily as spatial channel  \\
    \addlinespace
    Communication  (C)   
                               &  C1  &  Visualization interaction temporally needs to accommodate the handoff process  \\
                               &  C2  & Discussion of relevant time ranges  \\
                               &  C3  &  Comfortable with data sparsity  \\
                               &  C4  &  Comfortable with disparity between clinical conclusion and model result \\
                               &  C5  &  Impacts actionability \\
                               &  C6  &  Impacts trust \\
                               &  C7  &  Trend of data trumps singular predicted value \\
    \addlinespace
    Feature Descriptions (F)
                                      &  F1  & Related lab results  \\
                                      &  F2  &  List recent operations, interventions \\
                                      &  F3  &  Prioritization list  \\
                                      &  F4  &  Patient baseline \\
                                      &  F5  & Alert / alarm \\
\bottomrule
\end{tabular}
\end{table}

\end{document}